\documentstyle[12pt]{article}     
\newcommand{\rhohat}{{\pi_1}}
\begin{document}      
\begin{flushright}    
\small{hep-ph/0111361} \\         
\small{LA-UR-01-6434} 
\end{flushright}      
\vspace{1cm}          
       
\centerline{\Large\bf Molecular States and $1^{-+}$ Exotic Mesons}   
\vspace{1cm}          
       
\centerline{Rui Zhang$^{1,2,3}$, Yi-Bing Ding$^{1,4}$, Xue-Qian  
Li$^{1,2}$ and Philip R. Page$^{5,6,7}$}
       
\vspace{0.5cm}        
       
\noindent 1. CCAST (World Laboratory), P.O. Box 8730, Beijing 100080,      
China.\vspace{.2cm}              
       
\noindent 2. Department of Physics, Nankai University, Tianjin, 300071,    
China.\vspace{0.2cm}              
       
\noindent 3. Department of Physics, University of Minnesota, Minneapolis,  
MN 55455, U.S.A.\vspace{0.2cm}    
       
\noindent 4. Department of Physics, Graduate School, Academy of Sciences of
China, Beijing 100039, China. \vspace{0.2cm}      
       
\noindent 5. Theoretical Division, MS B283, Los Alamos National Laboratory, 
Los Alamos, NM 87545, U.S.A. \vspace{0.2cm}     
       
\noindent 6. Theoretical and Computational Physics Section, 
Oak Ridge National Laboratory, Oak Ridge, TN 37831-6373, U.S.A.\vspace{0.2cm}

\noindent 7. Department of Physics and Astronomy,
University of Tennessee, Knoxville, TN 37996-1501, U.S.A.

\vspace{1cm}          
       
\begin{center}        
\begin{minipage}{11cm}
\noindent {\bf Abstract}\\  
       
This work investigates whether the observed $1^{-+}$ exotic mesons are
molecular states. We first use a potential model to calculate the
spectra and lifetimes of the $f_0(980)$ and $a_0(980)$, taken to be
loosely bound molecular states of $K\overline K$, then apply the 
same scenario to the 
$1^{-+}$ exotic states $\rhohat(1400)$ and $\rhohat(1600)$, assuming them  
to be $\pi\eta(1295)$ and $\pi\eta(1440)$ molecules respectively.
We derive the effective potential in the framework of  
field theory at the hadronic level. Our results indicate that the
present data on $\rhohat(1400)$ and $\rhohat(1600)$ rule out the 
specific molecular ansatz.
We show that the lifetime of a loosely bound heavy--light
molecule with enough 
angular momentum is fully determined by the lifetimes of its
constituent mesons.
  
\end{minipage}
\end{center}
  
\vspace{2cm}
  
\baselineskip 22pt 
  
\section{Introduction}
  
\vspace{0.2cm}
  
New gluonic matter, such as glueballs and hybrids, are 
predicted by the general theory of QCD. But after more than 
three decades of effort, gluonic excitations 
have not been discovered unambiguously. 
%For low--lying glueballs, this may be due to large mixing between
%glueballs and mesons which makes identifying glueballs
%very difficult. 
The discovery of the $J^{PC}=1^{-+}$ isovector
resonances $\rhohat(1400)$ and $\rhohat(1600)$ is great progress along these 
lines, because it may imply that new gluonic matter 
has eventually been observed. The E852 Collaboration at BNL
observed $\rhohat(1400)$ in $\eta\pi$ in the 
reaction $\pi^-p\rightarrow \eta\pi^-p$,
with a mass and width of $1370\pm 16^{+50}_{-30}$ MeV and
$385\pm 40^{+65}_{-105}$ MeV respectively~\cite{E852}. 
Independently, $\rhohat(1400)$ has been observed 
in $\eta\pi$ with $M=1400\pm 20\pm 20$ MeV and 
$\Gamma = 310\pm 50^{+50}_{-30}$ MeV~\cite{Crys}, 
or $M=1360\pm 25\pm 20$ MeV and $\Gamma = 220\pm 90$
MeV~\cite{Crysa} by the Crystal Barrel Collaboration. The other  
resonance $\rhohat(1600)$ has been observed by E852 in $\rho\pi$ in 
$\pi^-p\rightarrow\pi^+\pi^-\pi^- p$ with mass 
$1593\pm8^{+29}_{-47}$ MeV and a width of 
$168\pm 20^{+150}_{-12}$ MeV~\cite{E852N}; in $\eta'\pi$
in $\pi^-p\rightarrow\eta'\pi^- p$ with
$M=1597\pm 10^{+45}_{-10}$ MeV and $\Gamma = 340\pm 40\pm 50$ 
MeV~\cite{E852a}; and most recently in a preliminary analysis in $b_1\pi$ in
$\pi^-p\rightarrow\omega\pi^-\pi^0p$ with
$M=1582\pm 10\pm 20$ MeV and $\Gamma = 289\pm 16\pm 27$~\cite{popov}.
The $\pi_1(1600)$ is observed decaying to the same final states
by VES~\cite{ves}.
  
A $1^{-+}$ meson cannot be a meson ($q\bar q$), a gluinoball ($\tilde{g}
\tilde{g}$) or a squarkball ($\tilde{q}\bar{\tilde{q}}$), 
because it is $J^{PC}$ exotic~\cite{squark}. 
The observed $1^{-+}$ mesons therefore
should be considered as the hybrid or glueball candidates which are
predicted by QCD. Since glueballs are isoscalar this possibility is excluded.
The $\rhohat(1600)$ is consistent with expectations for a hybrid 
meson~\cite{safir,Page}, while the $\rhohat(1400)$ is not~\cite{safir,ssr}.
Discovery of a hybrid is exciting, but  
identifying the observed $1^{-+}$ as the long--expected gluon excited states 
needs careful further study. Possible alternative hypotheses about their
structure have been discussed~\cite{Page}. There was an attempt to explain
the $\rhohat(1400)$ as a coupled channel effect dynamically generated in
$\pi\eta$ and $\pi\eta'$ scattering~\cite{bass}.
Donnachie and one of us (Page) 
suggested that the peak at 1400 MeV is a consequence of the 
interference of a 1600 MeV resonance with a non--resonant
Deck--type background with appropriate phase~\cite{Page1}.  
So far, this scenario has only been argued to account for the E852 data,
and has not been extended to the Crystal Barrel data. Also, aspects
of the proposed mechanism have been challenged on theoretical grounds  
\cite{ssr,tb}. However, it has been argued that there is no need 
for more than one resonance in the 1.3--1.6 GeV mass region 
when $\eta\pi,\; \eta^{'}\pi$ and $\rho\pi$ data are considered \cite{ach}. 
%The mass 
%spectrum and decay width of the exotic state is evaluated in the 
%color-flux-tube model where the hybrid structure is 
%supposed~\cite{Barnes,Isgur}.  
  
Before one can conclude that the observed $1^{-+}$ resonances are
gluon excited states,
even more possibilities should be explored. Among the possible  
alternatives, we explore the hypothesis of  
molecular structure. For $1^{-+}$ quantum numbers there may be many
plausible meson--meson bound states and their mixtures.
Since the binding energy for molecular  
states is not very large, the sum of the masses of the constituent 
mesons should be close to the mass of the exotic state.
According to the Particle Data Group~\cite{Data}, one of the favorable
combinations are that $\rhohat(1400)$ is composed of $\pi\; (0^{-+})$  
and $\eta(1295)\; (0^{-+})$ with total mass
$m_{\pi}+m_{\eta(1295)}\approx 1430\sim 
1438$ MeV, and $\rhohat(1600)$ is composed of $\pi$ and $\eta(1440)$ 
with total mass $m_{\pi}+m_{\eta(1440)}\approx 1535\sim 1620$ MeV.
The constituent mesons in these molecules have one unit of relative
angular momentum (``P--wave''). Since
$\Gamma_{\eta(1295)}=53\pm 6$ MeV and $\Gamma_{\eta(1440)}=50\sim 80$ 
MeV are not too wide, the $\eta(1295)$ and $\eta(1440)$ build
observable bound states with the pion, i.e. they can be constituents of  
molecular states with total widths of a few hundred MeV. The
possibility has also been discussed elsewhere~\cite{Godfrey}. 
In this work, we 
investigate whether the present data on $\rhohat(1400)$ and $\rhohat(1600)$
support this specific molecular hypothesis. 
Other possibilities for $\rhohat(1400)$, not discussed in this work, are a
$b_1$ and $\pi$ with no angular momentum between them (``S--wave''), denoted 
$(b_1\pi)_S$. Other S-- and P-- wave molecular possibilities are
$(f_1\pi)_S$ and $(K^{\ast} K)_P$. For $\rhohat(1600)$ the possibilities
are $(f_1(1420)\pi)_S,\; (f_1(1510)\pi)_S,\; (\rho\omega)_P$ and 
$(\rho(1450)\pi)_P$.  
  
We use a potential model to calculate the spectra and   
lifetimes of $\rhohat(1400)$ and $\rhohat(1600)$,
and then compare them with data to determine whether the molecular 
hypothesis is reasonable. We derive the potential between the 
constituent mesons by writing down  
the elastic scattering amplitude of the mesons in momentum  
space and then carrying out a Fourier transformation with respect to  
the 3--momentum of the exchanged mesons 
to obtain the potential in configuration space~\cite{Landau}. 
Substituting the potential into the 
Schr\"{o}dinger equation, we can obtain the binding energy and
wavefunction at the origin (or its slope) of the bound molecular
state. The effective couplings are obtained from 
data by assuming SU(3) flavour symmetry. To account for  
off--shell effects in the calculation of the amplitude, certain form  
factors are introduced. We write down the inelastic scattering 
amplitude of the mesons in momentum space, from which the lifetime
is calculated.
  
To test validity of the mechanism employed, we first apply it to 
study $f_0(980)$ and $a_0(980)$, which, in some interpretations, 
are interpreted as molecular states~\cite{Wein}. The molecular possibility 
has some support from lattice QCD~\cite{Stew}. 
As discussed above, we derive the potential between $K$ and 
$\overline K$, calculate the spectra of $f_0(980)$ and 
$a_0(980)$, obtain all
necessary parameters and then calculate their lifetimes. The  
results are in good agreement with data. Thus we are convinced
that the approach we adopt is applicable to $1^{-+}$ if it is a  
molecular state. 
  
The paper is organized as follows. After this introduction, we
present most of the formulation and some technical details in Section 2.
In Section 3, we report our numerical results and the last section 
is devoted to our discussion and conclusion. 
  
\section{Formulation} 
  
\subsection{Exchange mesons\label{sec1}}  
  
Consideration of composites of two quarks and two antiquarks is  
complicated by the fact that different colour neutral combinations 
are possible. The first possibility is that no subcomponents of the
system is colour neutral, commonly referred to as a four--quark state.
The second possibility is that one quark--antiquark subcomponent is
colour neutral, forming a meson, and that the same happens to the
remaining subcomponent. This composite of two mesons is referred to as a 
molecule. The latter possibility 
will be the subject of this work. In the limit of a large number of
colours $N_c$ in QCD there are no four--quark states, but only 
meson--meson molecules~\cite{witten}.
The molecular picture becomes more  
accurate if the mesons contained in the molecule are separated far from 
each other relative to the size of the constituent mesons. This would 
happen when the binding energy of the two mesons is small relative to 
the quark--antiquark energy within each individual meson
(which approximately equals the mass of the meson when the quarks
are light).
The theoretical calculation of such a small binding energy of mesons  
is complicated because the binding energy is the result of delicate
cancellations between attractive and repulsive interactions, each of which 
individually corresponds to much larger energies than the total binding 
energy. This follows because individual interactions are governed by  
the QCD scale $\Lambda_{QCD}$, which is much larger than the binding energy. 
The calculation of the binding energy has been done in lattice 
QCD~\cite{Stew,lat},
QCD sum rules~\cite{qcdsr},
quark--antiquark pair exchange models~\cite{tb,dooley},
four--quark potential~\cite{Wein,semay}, flux--tube~\cite{carlson} 
and other~\cite{other} models, 
and molecular models with meson exchange~\cite{Tor}. We adopt the latter.
Meson exchange models should be most accurate when the binding energy of
the constituents is low. This is because in this case the constituents
are typically separated far apart, so that the cost that confinement
imposes on exchanging coloured objects at large distances
means that the exchange is more likely
to be a colour singlet, i.e. a meson.   
Consideration of a system 
with low binding energy, i.e. of two mesons 
separated far apart, yields the result that exchange mesons with  
a long range would be most important for the calculation of the binding 
energy. 
Long range exchange mesons correspond to low mass $m_q$ exchange mesons,
taken to be either the $\pi$, $\sigma$ (also called the  
$f_0(400-1200)$), $\eta$, $\rho$, $\omega$, $\eta'$, $f_0(980)$,
$a_0(980)$ or $\phi$ in this work. An exchange meson can be neglected
when its range $\sim \frac{1}{m_q}$ is small with respect to the r.m.s. radius
$r_{rms}$ of the molecular state, i.e. when $1\over m_q r_{rms}$ is small.

The general formalism for meson exchange is reviewed in 
Appendix \ref{app}. For the remainder of this section we describe
the theoretical formulation used in our calculations.

Because the momentum of the exchanged meson in t--channel is 
space--like while it is time--like in s--channel, the propagator is 
expressed as
\begin{eqnarray} 
{1\over q^2-m_q^2} && \;\;\;\; {\rm t-channel}\\
{1\over q^2-m_q^2+i\Gamma_mm_q} &&\;\;\;\;{\rm s-channel},  
\end{eqnarray}
where the Breit--Wigner form is taken and $\Gamma_m$ is the
total width of the exchanged meson. At vertices, we use the  
standard strong coupling forms~\cite{Chung} and the proper  
Clebsch-Gordon coefficients to manifest the isospin structure
of
the vertices. The couplings take the values extracted from  
corresponding decay widths.   
There are three types of decay modes relevant to this work: 
$$0^{+}\rightarrow 0^-0^- \;({\rm S-wave}),\;\;\; 1^-\rightarrow 
0^-0^-\;({\rm P-wave}),\;\;\; 1^-\rightarrow 0^-1^-\;({\rm  
P-wave}).$$ 
The decay widths are  
\begin{equation}\label{wid1}
\Gamma={l\over(2 J+1)}{1\over 8\pi m^2}|M|^2,
\end{equation}
where $l$ is the magnitude of the 
3--momentum of the decay product in the CM
frame, $m$ and $J$ are the mass and internal angular momentum 
of the decaying meson, and $M$ is the amplitude, with  
\begin{eqnarray}\label{amp} 
(i) && |M|^2=g_1^2,\;\hspace{6.85cm} {\rm for\; 0^+\rightarrow 0^-0^-},
\nonumber\\ 
(ii) &&
|M|^2=g_2^2(p_1-p_2)^{\mu}(p_1-p_2)^{\nu}\sum_{\lambda}
\epsilon_{\mu}^{\lambda*} 
\epsilon_{\nu}^{\lambda},\;\hspace{1.78cm} {\rm for\; 1^-\rightarrow 0^-0^-},  
\nonumber \\
(iii) && |M|^2=g_3^2\epsilon_{\alpha\beta\mu\nu}  
\epsilon_{\alpha'\beta'\mu'\nu'} 
P^{\alpha}P^{\alpha'}(p_1-p_2)^{\mu}(p_1-p_2)^{\mu'} \nonumber \\
&& \times\sum_{\lambda_1}\epsilon^{\beta}_{\lambda_1}  
\epsilon^{\beta'*}_{\lambda_1} 
\sum_{\lambda_2}\epsilon^{\nu}_{\lambda_2}\epsilon^{\nu'*}_{\lambda_2}\;
\hspace{5.0cm} {\rm for\; 1^-\rightarrow 0^-1^-}, 
\end{eqnarray}
where $g_i$ is the corresponding strong coupling, $P$ is 
the 4--momentum of the decaying meson, $p_1,p_2$ are the momenta of 
the outgoing mesons, and $\epsilon_{\lambda}$ is the polarization vector of the
vector meson. We note that all the mesons which are connected to the  
vertex are on their mass shells. Using Eqs. \ref{amp},  
the coupling constant $g_i$ is calculated from the experimental width.
If the experimental width is unavailable, $g_i$
is obtained by assuming SU(3) flavour symmetry for the
coupling.
  
Since the propagators are off--shell, we introduce form factors 
for the effective couplings $g_i$ at the vertices. The usual form 
factors are expressed as~\cite{Tor,Li}  
\begin{equation}{\Lambda^2-m_q^2\over \Lambda^2-q^2},\end{equation}
at each vertex with $\Lambda$ is an adjustable constant which models the
off--shell effects at the vertices due to the 
internal structure of the mesons. 
%In fact, these form factors can partially
%represent the QCD Van der Waals 
%effects which are induced by the 
%finite sizes and internal structure of the constituent mesons.
%Its effects correspond to 
%the Van der Waals force in the thermodynamics of a molecule in a gas. 
 
\subsection{Effective potential between two constituent  
mesons\label{sec2}}
  
Following the standard procedure given in ref.~\cite{Landau} and using
the effective vertices, we can write down the scattering 
amplitude between the two constituent mesons in momentum space  
and convert it into configuration space by a Fourier transformation.
This procedure was used to 
derive the effective potential between a quark and an antiquark~\cite{tb}. 
We consider all pseudoscalar, 
scalar and vector meson exchanges in t--, s-- and
u--channel, and indicate explicit expressions for the processes  
that need to be included in our calculations.
In the center of mass (CM) frame of the bound state, with
  
{\it (i)} $0^+$ the exchanged meson in t--channel,  
\begin{equation}\label{pot1}               
V({\bf q})=-\frac{g_1^2}{{\bf q}^{2}+m_{q}^{2}}                     
\left(\frac{\Lambda^{2}-m_{q}^{2}}{\Lambda ^{2}+{\bf q}^{2}}\right)^{2}       
\frac{1}{4m_{1}m_{2}}\left[\left(1-\frac{{\bf p}^{2}}{2m_{1}^{2}}-
\frac{{\bf p}^{2}}{2m_{2}^{2}}\right) 
-(2{\bf p}\cdot{\bf q}-{\bf q}^2)({1\over 4m_1^2}+{1\over 4m_2^2}) \right],
\end{equation}                                                                
  
{\it (ii)} $1^-$ the exchanged meson in t--channel, 
\begin{equation} 
V({\bf q})=\frac{g_2^2}{{\bf q}^{2}+m_{q}^{2}} 
\left(\frac{\Lambda^{2}-m_{q}^{2}}{\Lambda ^{2}+{\bf q}^{2}}\right)^2   
(1+\frac{{\bf p}^2}{m_1 m_2}-{{\bf p}\cdot{\bf q}\over m_1m_2} 
+\frac{{\bf q}^2}{4m_1 m_2}),                         
\end{equation}
  
{\it (iii)} $0^+$ the exchanged meson in s--channel,
\begin{eqnarray}\label{pot3}
V({\bf q}) & = & \frac{g_1^2(\Lambda^2 -m_q^2)^2}{4m_1 m_2}
\left[\frac{1}{(a^{\prime 2}-m_q^2)(a^2 -\Lambda^2)^2} 
-\frac{b^{\prime}{\bf p}^2}{(a^{\prime 2}-m_q^2)^2
(a^2 -\Lambda^2)^2}\right.\nonumber\\
&& \left.-\frac{2b{\bf p}^2}{(a^{\prime 2}-m_q^2)(a^2 -\Lambda^2)^3}  
-\frac{(b^2-2b){\bf p}^2} 
{2a^2(a^{\prime 2}-m_q^2)(a^2 -\Lambda^2)^2}\right],
\end{eqnarray}
where we introduce the variables
\begin{eqnarray} 
a^2 & = & (m_1+m_2)^2\\
b & = & \frac{(m_1 +m_2)^2}{m_1 m_2}\\  
a^{\prime 2} & = & a^2+\frac{\Gamma_q^2 m_q^2}{a^2 -m_q^2}\\
b^{\prime} & = & b\left(1-\frac{\Gamma_q^2 m_q^2}{(a^2 -m_q^2)^2} \right
),
\end{eqnarray}
where ${\bf q}$ is the 3--momentum that is exchanged.
The momenta of constituent mesons with mass $m_1$ or $m_2$
in the initial state is respectively denoted by ${\bf p}$ and $-{\bf p}$.
After the Fourier transformation, we obtain the potential forms in 
configuration space:   
\begin{eqnarray} 
&&\hspace{-3.65cm}(i)\; 0^+\; \mbox{in t--channel},  \nonumber\\ 
 V({\bf r}) &=& -\frac{g_1^2}{16\pi m_1 m_2}\left [            
\frac{e^{-m_q r}-e^{-\Lambda r}}{r}                
+\frac{(m_q^2-\Lambda^2)e^{-\Lambda r}}{2\Lambda}\right.\nonumber\\  
&& \left.-\frac{m_1^2+m_2^2}{2m_1^2 m_2^2}\                       
\frac{e^{-m_q r}-e^{-\Lambda r}}{r}\hat{{\bf p}}^2          
+\frac{m_1^2+m_2^2}{2m_1^2 m_2^2}\                       
\frac{(m_q^2-\Lambda^2)e^{-\Lambda r}}{2\Lambda}\hat{{\bf p}}^2 
\right ]\nonumber \\                           
&& +{ig_1^2(\Lambda^2-m_q^2)^2\over 2m_1m_2}\left[  
A(m_qr+1)e^{-m_qr}+B(\Lambda r+1)e^{-\Lambda r}\right.\nonumber \\  
&& \left. +(C-B\Lambda^2)e^{-\Lambda r}r^2\right]            
\left({1\over 4m_1^2}                                     
+{1\over 4m_2^2}\right){\hat{\bf p} \cdot{\bf r}\over 4\pi r^3}\nonumber\\ 
&& -{g_1^2(\Lambda^2-m_q^2)^2\over 4m_1m_2}     
\left[{e^{-\Lambda r}\over\Lambda}+Am_q^2{e^{-m_q r}\over r} 
+Bm_q^2 {e^{-\Lambda r}\over r}\right.\nonumber\\           
&& \left.+(C-B\lambda^2)m_q^2{e^{-\Lambda                   
r}\over\Lambda}\right]\left({1\over 4m_1^2}             
+{1\over 4m_2^2}\right),                           
\end{eqnarray}                                
where $\hat{\bf p}$ is the momentum operator and              
$$A={1\over(\Lambda^2-m_q^2)^2},\;\; B=-A,\;\; C={m_q^2-2\Lambda^2\over   
(\Lambda^2-m_q^2)^2},$$      
\begin{eqnarray}                     
&&\hspace{-2cm}(ii)\; 1^-\; \mbox{in t--channel},  \nonumber\\  
V({\bf r}) &=& \frac{g_2^2}{4\pi}\left[                   
\frac{e^{-m_q r}-e^{-\Lambda r}}{r}                       
+\frac{(m_q^2-\Lambda^2)e^{-\Lambda r}}{2\Lambda}\right.\nonumber\\ 
&& \left. +\frac{1}{4m_1 m_2}\frac{m_q^2(e^{-m_q r}-e^{-\Lambda r})}{r} 
+\frac{1}{4m_1 m_2}\frac{(m_q^2-\Lambda^2)\Lambda e^{-\Lambda     
r}}{2}\right.                                                    
\nonumber\\                                                       
&& +\left.\frac{1}{m_1 m_2}\frac{e^{-m_q r}-e^{-\Lambda r}}{r}\hat{{\bf p}}^2
+\frac{1}{m_1 m_2}\frac{(m_q^2-\Lambda^2)e^{-\Lambda r}}{2\Lambda}      
\hat{{\bf p}}^2                                                         
\right ]\nonumber\\                                                      
&& -{ig_2^2(\Lambda^2-m_q^2)^2\over 4\pi m_1m_2}                         
\left[A(m_qr+1)e^{-m_qr}+B(\Lambda r+1)e^{-\Lambda r} \right.\nonumber\\  
&& \left. +(C-B\Lambda^2)e^{-\Lambda r}r^2\right]{
\hat{\bf p}\cdot{\bf r}\over 
4\pi r^3},\\                                                                
&&\hspace{-2cm}(iii)\; 0^+\; \mbox{in s--channel},  \nonumber\\ 
V({\bf r}) &=& \frac{g_1^2(\Lambda^2 -m_q^2)^2}{4m_1 m_2}  
\left[\frac{1}{(a^{\prime 2}-m_q^2)(a^2 -\Lambda^2)^2} 
-\frac{b^{\prime}\hat{{\bf p}}^2}{(a^{\prime 2}-m_q^2)^2(a^2 -\Lambda^2)^2}
\right.\nonumber\\ 
&& \left.-\frac{2b\hat{{\bf p}}^2}{(a^{\prime 2}-m_q^2)(a^2 -\Lambda^2)^3} 
-\frac{(b^{2}-2b)\hat{{\bf p}}^{2}}{2a^2(a^{\prime 2}-m_q^2)
(a^2 -\Lambda^2)^2}
\right ]\delta^3({\bf r})
,\\ 
&&\hspace{-2cm}(iv)\; 0^+\; \mbox{in u--channel},   \nonumber\\
V({\bf r}) &=& -\frac{g_1^2(\Lambda^2 -m_q^2)^2}{16\pi rm_1 m_2}
\left\{\left[\frac{re^{-dr}}{2d(c^2+d^2)}+\frac{e^{-dr}}{(c^2+d^2)^2} 
+\frac{\cos (cr)}{(c^2+d^2)^2} \right ]\right.\nonumber\\
&& -\left [\frac{re^{-dr}}{8d^3(c^2+d^2)}+\frac{r^2e^{-dr}}{8d^2(c^2+d^2)} 
+\frac{re^{-dr}}{2d(c^2+d^2)^2}+\frac{e^{-dr}}{(c^2+d^2)^3} 
-\frac{\cos (cr)}{(c^2+d^2)^3} 
\right ]2b_u\hat{{\bf p}}^2 \nonumber\\ 
&& -\left [\frac{r\sin (cr)}{2c(c^2+d^2)^2}+\frac{2\cos (cr)}{(c^2+d^2)^3} 
+\frac{re^{-dr}}{4d(c^2+d^2)^2}+\frac{e^{-dr}}{(c^2+d^2)^3} \right ]b_u 
\hat{{\bf p}}^2\nonumber\\
&& -\left.\left [\frac{re^{-dr}}{2d(c^2+d^2)}+\frac{e^{-dr}}{(c^2+d^2)^2}  
+\frac{\cos (cr)}{(c^2+d^2)^2} 
\right ]\frac{(b_u^2-2b_u)}{2a^2}\hat{{\bf p}}^2  
\right \}
,
\end{eqnarray}
where the last potential is only displayed in position space for brevity and 
some small relativistic contributions to it are neglected. Here
\begin{equation} 
a_u^2=(m_2-m_1)^2,\;\; b_u={-(m_2-m_1)^2\over m_1m_2},     
\end{equation} 
and 
\begin{equation} 
c^2=a_u^2-m_q^2=(m_2-m_1)^2-m_q^2,\;\;    
d^2=\Lambda^2-a_u^2=\Lambda^2-(m_2-m_1)^2.
\end{equation}

Since the mesons are colour singlets, we do not need to introduce a 
``confinement" potential as for the quark--antiquark system.  
Substituting the potentials into the Schr\"{o}dinger equation, 
we can solve it and obtain the eigenenergy and wavefunction. As 
noted previously, our interest is in
molecular states with small binding energy. In such a molecule the 
kinetic energy of the constituent mesons in the bound state is
also small, so that relativistic effects can be neglected,
and the non--relativistic Schr\"{o}dinger equation can be used.  
  
\subsection{Decay width\label{decwid}}
  
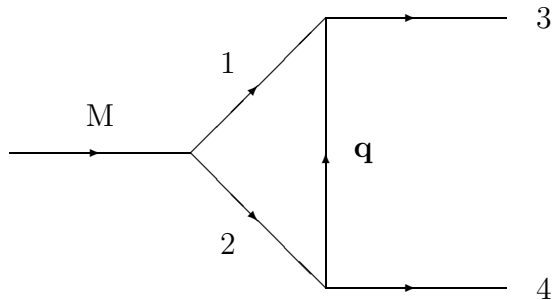
\begin{figure}[ht]
\unitlength=1.00mm 
\special{em:linewidth 0.4pt}
\linethickness{0.4pt} 
\begin{picture}(129.00,50.00)(0,60) 
\put(38.00,82.00){\vector(1,0){12.00}}  
\put(50.00,82.00){\line(1,0){12.00}}
\put(62.00,82.00){\vector(1,1){9.00}}
\put(71.00,91.00){\line(1,1){9.00}} 
\put(62.00,82.00){\vector(1,-1){9.00}}  
\put(71.00,73.00){\line(1,-1){9.00}}
\put(67.00,94.00){\makebox(0,0)[cc]{1}} 
\put(67.00,70.00){\makebox(0,0)[cc]{2}} 
\put(50.00,87.00){\makebox(0,0)[cc]{M}} 
\put(80.00,64.00){\vector(0,1){18.00}}  
\put(80.00,82.00){\line(0,1){18.00}}
\put(80.00,100.00){\vector(1,0){12.00}} 
\put(92.00,100.00){\line(1,0){12.00}}
\put(80.00,64.00){\vector(1,0){12.00}}  
\put(92.00,64.00){\line(1,0){12.00}}
\put(109.00,100.00){\makebox(0,0)[cc]{3}} 
\put(109.00,64.00){\makebox(0,0)[cc]{4}}
\put(85.00,82.00){\makebox(0,0)[cc]{\bf q}} 
\end{picture} 
\caption{The decay of meson M. Mesons 1 and 2 are the 
constituent mesons in the molecular state. Mesons 3 and 4  
are the decay products.}  
\end{figure}
  
The decay occurs via the triangle diagram shown in Fig. 1.
The case of t--channel meson exchange is displayed. The
s--channel is similar.
The diagram represents a Bethe--Salpeter (B--S) equation where
the internal momentum in the loop is integrated out.  
The two intermediate lines labelled ``1'' and ``2''
represent the constituent  
mesons. In a loosely bound molecule
they are very close to their mass shell. Thus,
instead of the complicated B--S integration, we treat the
two constituent mesons as free ones, and they transit  
into the final states via an inelastic scattering 
process. 
  
When decay via
%t--channel 
meson exchange takes place, the
amplitude is approximately proportional to the wavefunction (or  
its derivative) at the origin. This is due to an integration effect.
When we take the B--S approach to calculate the amplitude via a 
loop containing the bound state with an appropriate kernel (the  
potential we derived above), we immediately find that after 
integrating out the internal momentum, only a wavefunction at 
origin $\psi(0)$ remains when the constituent mesons are in
relative S--wave. From 
the formula $|\psi(0)|^2=2\mu<\psi|{dV\over dr}|\psi>$~\cite{Quigg}, 
with $\mu$ the reduced mass of the system,
it follows that $\psi(0)$ is an average over the entire $r-$region, 
in other words, effects of all ranges in $r$ are
involved in $\psi(0)$. So even though only the wavefunction at the
origin matters, this does not mean that only small distance effects
are important.
For constituent mesons in P--wave we know that the coupling 
is proportional to the momentum. After integration in the B--S equation,
it turns into the derivative of the wavefunction at origin $\psi'(0)$. 
  
Let us briefly discuss the legitimacy of the treatment adopted in 
the decay calculation.
%This approach is based on the non--relativistic approximation, i.e. 
%the relative momentum of the constituent mesons is small compared
%to the masses of the constituents. For molecular states, this is 
%the case. We can further argue that as long as the exchanged meson 
%is heavy enough compared to the momentum which should be integrated over,  
%the whole picture would make more sense because the two constituents 
%are nearly 
%on their mass shell. In fact, in the B--S integration,
%$(|{\bf p}-{\bf q}|)/m_q$ is the relevant variable and the whole
%expression can be expanded with respect to it. As long as it is  
%small, the leading term in approximation is accurate enough and  
%this term results in the wavefunction at origin (in other words, 
%the higher orders in momentum can be safely ignored).  
%Thus since the pole position in the B--S integration determines 
%$|{\bf q}|$ to be small,  
We find that the validity of the approximation used in the 
B--S approach depends on the size of 
${|{\bf p}|\over m_q}$ being small. From the
Heisenberg uncertainty principle, $|{\bf p}|\sim \frac{1}{r_{rms}}$.
%where $r_{rms}$ is the rms radius of the molecular state and the ratio 
%{|{\bf p}|\over m_q}=1/m_qr_{rms}$ being small is the condition 
%for using the approximation. 
For a molecular  
state with small binding energy $E_b$, the
r.m.s. radius $r_{rms}$ is relatively large ($E_b\sim 1/(2\mu r^2_{r.m.s.})$).
%, the exchanged  
%mesons must be at least $\rho (1^{-})$ or $\sigma(0^+)$ as  
%aforementioned,  
Thus ${|{\bf p}|\over m_q}$ is small as expected.

%In this approximation, we can take the three--momentum of the 
%intermediate mesons to be zero, $|{\bf p_1}|=|{\bf p_2}|=0$ 
%and $E_1\sim m_1,\;E_2\sim m_2$. 
The decay can be realized  
via t-- and s--channels and their contributions must be
summed up. For $0^-0^-\rightarrow 0^-0^-$ scattering, exchanging
  
$(i)\; 0^+$ in t--channel, 
  
\begin{eqnarray} 
{\cal M}_t=\frac{g_1^2}{q^{2}-m_{q}^{2}} 
\left(\frac{\Lambda ^{2}-m_{q}^{2}}{\Lambda ^{2}-q^{2}}\right)^{2}, 
\end{eqnarray}
  
$(ii)\; 1^-$ in t--channel,
\begin{eqnarray} 
{\cal M}_t&=
&g_2^2\left(\frac{\Lambda ^{2}-m_{q}^{2}}
{\Lambda ^{2}-q^{2}}\right)^{2}
(p_1+p_3)^{\mu}(p_2+p_4)^{\nu}D_{\mu\nu}(q)\nonumber\\&= 
& -\frac{gg^{\prime}}{q^2-m_q^2} 
\left(\frac{\Lambda ^{2}-m_{q}^{2}}{\Lambda^{2}-q^{2}}\right)^{2}\nonumber\\ 
&& \left [\frac{2M^2-m_1^2-m_2^2-m_3^2-m_4^2}{2}  
+m_1\sqrt{m_4^2+l^2}+m_2\sqrt{m_3^2+l^2}\right.\nonumber\\  
&& \left.+\frac{(m_1^2-m_3^2)(m_2^2-m_4^2)}{m_q^2}\right ],  
\end{eqnarray}
  
$(iii)\; 0^+$ in s--channel,  
\begin{eqnarray} 
{\cal M}_s=\frac{g_1^2(q^2-m_q^2-i\Gamma_qm_q)}{(q^{2}-m_{q}^{2})^2+ 
\Gamma_q^2m_q^2} 
\left(\frac{\Lambda ^{2}-m_{q}^{2}}{\Lambda ^{2}-q^{2}}\right)^{2}, 
\end{eqnarray}
where $m_3$ and $m_4$ are the masses of the outgoing mesons in Fig. 1.
Besides these three amplitudes, there are those corresponding 
to $0^-0^-\rightarrow 0^-1^-$. We omit their explicit expressions. 
  
The decay width is 
\begin{eqnarray} \label{swav}
(i) \;\; \Gamma_{0^+}={|\psi(0)|^2l\over 16\pi 
m^3}\;|\sum_{i=t,s,u}{\cal M}_i|^2 \;\;\;\;&& {\rm 
for}\;0^+\; {\rm decay},\\  \label{pwav}
(ii)\;\;\Gamma_{1^-}={|\psi'(0)|^2l\over 48\pi 
m^5}\;|\sum_{i=t,s,u}{\cal M}_i|^2 \;\;\;\;&& {\rm 
for}\;1^-\; {\rm decay},
\end{eqnarray}
where $i$ is summed over the s--, t-- and u--channel contributions.
$l=(m^4+m_3^4+m_4^4-2m^2m_3^2-2m^2m_4^2-2m_3^2m_4^2)^{1/2}/(2m)$ 
is the 3--momentum of the outgoing mesons in the CM frame.
  
\section{Numerical results} 
  
Numerical results are obtained by including the $\sigma$ 
 as an exchanged scalar meson, and the
$\omega, \; \rho$ and $\phi$ as exchanged vector mesons. 
There is no low--lying isovector scalar meson, and the exchange of the
low--lying pseudoscalar mesons $\pi$, $\eta$ and $\eta'$ is forbidden 
by parity. Vector meson exchange in s--channel is not allowed by conservation
of charge conjugation.
For the $\rhohat(1400)$ and $\rhohat(1600)$ systems we also incorporated
$f_0(980)$ and $a_0(980)$ exchange.
%u--channel scattering is not considered, because pseudoscalar meson
%exchange is forbidden by parity and $\sigma$ meson exchange is 
%forbidden by isospin.
  
\subsection{$f_0(980)$ and $a_0(980)$}  
  
First we consider $f_0(980)$ and $a_0(980)$ which are  
taken as bound states of $K\overline K$. Since all other 
parameters are fixed by experimental data and SU(3) flavour symmetry, only
the off--shellness parameter $\Lambda$ is a free parameter. We do not 
intend to determine its value by 
fitting the mass spectra of $f_0(980)$ and $a_0(980)$.
In a reasonable range of 1 to 10 GeV, we choose 
a value of $\Lambda$ which gives reasonable predictions. We have 
$$\Lambda=3.4 \;{\rm GeV}.$$% 3.44 GeV 
The resultant masses are $$ M_{f_0(980)}=1013 \;{\rm
MeV}\;\;\;M_{a_0(980)}=996\;{\rm MeV}\;\;\;{\rm without\;   
zero-energy},$$  
$$ M_{f_0(980)}=999 \;{\rm
MeV}\;\;\;M_{a_0(980)}=982\;{\rm MeV}\;\;\;{\rm with\; zero-energy},$$ 
and should be compared to the experimental data~\cite{Data}  
$$ M_{f_0(980)}=980\pm 10 \;{\rm 
MeV}\;\;\;M_{a_0(980)}=984.8\pm 1.4\;{\rm MeV}.$$
The calculated values are fairly near to the experimental values,
given that $\Lambda$ is the only free parameter we have for the  
$K\overline K$ system.
There is always the possibility to add an additional constant term
to the potentials that we derived, called the ``zero--energy''. 
The term ultimately arises from renormalization when 
non--perturbative QCD is matched to a meson exchange model.
In our final mass estimate we included a zero--energy
so that the calculated values
are in closer agreement with experiment.
%In fact, in the heavy quarkonium case, for fitting data, usually a    
%zero-point energy is added to the eigenvalue obtained by solving
%the Sch\"{o}dinger equation and its physical meaning is similar to    
%the zero-point energy of ${1\over 2}\hbar\omega$ for the harmonic     
%oscillator in quantum mechanics, which cannot be obtained from the    
%equation. Here, it seems that we do not need to introduce the  
%non-perturbative confinement term by hand and the zero-point   
%energy is also not necessary. It makes more sense.
  
The central experimental value of
$M_{a_0(980)}$ is a bit larger than that of $M_{f_0(980)}$, but our
result is inverted. Our results may be changed as follows.  
The $a_0(980)$ receives no contribution from s--channel $\sigma$ exchange  
due to isospin conservation. However, the $f_0(980)$ does.  
In the case of the $f_0(980)$, the $\sigma$ intermediate meson
constructively interferes with t--channel exchange. 
However, obtaining coupling constants from widths, as we do, means 
that we cannot fix the sign of the amplitude. So there is the possibility  
that the amplitude has an additional $-1$ between the two
channels. This may well change the results. Moreover,  
if we choose different values of $\Lambda$ 
for $f_0(980)$ and $a_0(980)$ and fine--tune other
parameters, we can substantially improve our fit of the spectra. 
Since our aim is to obtain a general picture with the least parameters, 
we shall not pursue these further refinements. 
Because parameter changes allow considerable
change in binding energy, 
we do not view our calculated binding energy as accurate. 
Our concern is with an overall understanding of the spectra  
and lifetimes, in order to draw  
qualitative conclusions. The details and  
subtle deviations are not important. 

For both $f_0(980)$ and $a_0(980)$
$$|\psi(0)|^2=1.2 \times 10^{-3}\;{\rm GeV^3},$$  %0.00122  
showing that the states are loosely bound.
Substituting this value into Eq. \ref{swav}, we 
obtain 
\begin{eqnarray} 
\Gamma(f_0(980))\rightarrow\pi\pi) &=& 108\; {\rm MeV \;\;\;
\mbox{s-- and t--channels}\; interfere\; constructively},\\ 
\Gamma(f_0(980))\rightarrow\pi\pi) &=& 86.5\; {\rm MeV \;\; 
\mbox{s-- and t--channels}\; interfere\; destructively},\\  
\Gamma(a_0(980)\rightarrow\pi\eta) &=& 36.3\; {\rm MeV}.
\end{eqnarray}
We display values corresponding to the possibilities of the 
s-- and t--channels interfering without or with an additional 
sign between the two channels. 
  
The data indicate that the modes 
$f_0(980)\rightarrow \pi\pi$ and 
$a_0(980)\rightarrow\pi\eta$ are dominant, i.e., these 
partial widths can approximately be taken as the total widths.
The experimental data are 
$$\Gamma(f_0(980))=40\sim 100\;{\rm MeV}\;\;{\rm  
and}\;\; \Gamma(a_0(980))=50\sim 100\;{\rm MeV}.$$
The values we calculated were not fitted to this data, 
so that the agreement with experiment is noticible. 
  
\subsection{$\rhohat(1400)$ and $\rhohat(1600)$}  
  
We now calculate the spectra and the decay
widths of the $1^{-+}$ exotic states, assuming  them
to be molecular states.
As discussed above, we postulate that $\rhohat(1400)$ and
$\rhohat(1600)\; (1^{-+})$ are molecular states of  
$\pi\eta(1295)$ and $\pi\eta(1440)$. With the potential
derived in Section 2, we obtain the following results. 

\noindent For $\rhohat(1400)$
$$\Lambda=1.17\;{\rm GeV}, E_b=-78\;{\rm MeV},  %-0.07835 GeV 
M_{\pi\eta(1295)}=1354\;{\rm MeV},\; |\psi'(0)|^2=3.95\times
10^{-5}\;{\rm GeV^5}.$$
For $\rhohat(1600)$
$$\Lambda=1.35\;{\rm GeV}, E_b=-14\;{\rm MeV}, %-0.01438 GeV
M_{\pi\eta(1440)}=1521\sim 1591\;{\rm MeV},\;|\psi'(0)|^2=4.18\times  
10^{-5}\; {\rm  GeV^5}.$$
  
By contrast to the case for $f_0(980)$ and $a_0(980)$, there are 
several decay modes for the exotic states. We can only evaluate a
few exclusive modes and later by a general argument, we can obtain 
the order of magnitude of their lifetimes. Below we list several 
partial widths:
\begin{eqnarray} 
 \Gamma(\rhohat(1400)\rightarrow \pi\eta)& =& 5.28\times 10^{-2}\;{\rm
MeV},\nonumber \\
 \Gamma(\rhohat(1400)\rightarrow \pi\eta')& =& 6.80\times 10^{-2}\;{\rm 
MeV},\nonumber \\
 \Gamma(\rhohat(1400)\rightarrow \pi^-\rho^0)& =& 2.86\times 10^{-2}\;{\rm
MeV},\nonumber \\
 \Gamma(\rhohat(1600)\rightarrow \pi\eta)& =& 4.14\times 10^{-2}\;{\rm
MeV},\nonumber \\
 \Gamma(\rhohat(1600)\rightarrow \pi\eta')& =& 6.56\times 10^{-2}\;{\rm 
MeV},\nonumber \\
 \Gamma(\rhohat(1600)\rightarrow \pi^-\rho^0)& =& 3.70\times 10^{-2}\;{\rm 
MeV},  
\end{eqnarray}
where the calculation was performed for
$\rhohat_{(\pi^-\eta(1295))}(1400)$ and
$\rhohat_{(\pi^-\eta(1440))}(1600)$.
  
In our model, we can estimate the relative decay widths of  
$\rhohat(1600)$ to $\pi b_1,\pi f_1$ and $\pi\rho$. The difference is 
due to the effective couplings and the form factor parameter
$\Lambda$. With isospin symmetry, we have 
$$\pi b_1:\pi f_1:\pi\rho\approx 2.6:2:1.$$  
One hence obtains various branching ratios for 
different decay modes, but one can conclude that with the hypothesis 
that the $1^{-+}$ exotic states are molecular states,  
$\pi\eta,\;\pi\eta',\;\pi\rho,\;\pi b_1$ and 
$\pi f_1$ are of the same order of  
magnitude.  
  
\section{Discussion and conclusion} 
  
In this work we consider molecules of two constituent mesons with meson
exchange. We argued in section \ref{sec1} that the picture becomes
more accurate if the binding energy of the constituent mesons is small
compared to their masses. We also argued that
low mass exchanges dominate in a meson exchange model if the
binding energy is small compared to the exchange meson mass. In
section \ref{sec2} we argued that small binding energy enables the
use of a non--relativistic Schr\"{o}dinger equation. In section \ref{decwid}
we argued that the decay formalism is valid when the binding energy is
small with respect to the mass of the exchange meson. 
To summarize, our formalism should be valid when the binding energy
is small relative to the masses of the constituent and exchange meson
masses. The lowest mass constituent meson considered in this work is
the $\pi$, and the lowest mass meson exchanged is found to be the $\sigma$.

We investigate the possibility that the observed $1^{-+}$ 
exotic states $\rhohat(1400)$ and $\rhohat(1600)$ are molecular
states.
  
Our strategy is to use the potential derived from 
field theory to calculate the spectra and partial decay widths
and see if they can fit the measured data.
We derive the potential by
calculating the elastic scattering amplitude between the 
constituent mesons and then carrying out a Fourier transform to
convert the amplitude into the potential in configuration
space. In our calculations, we obtain the  
effective couplings from data by assuming SU(3) flavour symmetry. Only the  
free parameter $\Lambda$, which models the off--shell effects
at the effective vertices, needs to be determined. 
The inelastic scattering amplitude in momentum space is used to 
calculate the lifetimes.
  
To check the plausibility and validity of the approach, we study 
$f_0(980)$ and $a_0(980)$ which are commonly considered as  
molecular states of $K\overline K$.   
The mass spectra and lifetimes 
of $f_0(980)$ and $a_0(980)$  
are in qualitative agreement with data. Thus
we are convinced that the potential approach is applicable for
evaluating the approximate mass spectrum and lifetimes of molecular 
states.  
  
With the same scheme, we calculate the decay widths of the
observed $1^{-+}$ states $\rhohat(1400)$ and $\rhohat(1600)$. We assume 
them to be molecular states of $\pi\eta(1295)$ and
$\pi\eta(1440)$ respectively.  
%That is because in other structures, sum of the
%constituent meson masses may be too far from the observed masses 
%of $\rhohat(1400)$ and $\rhohat(1600)$ and one does not expect too large
%binding energy for molecular states.
By fitting the spectra of $\rhohat(1400)$ and $\rhohat(1600)$, we fix 
the $\Lambda$ parameter and obtain $|\psi'(0)|^2$ which plays a 
crucial role in the decay width calculation. 
  
The value of $|\psi'(0)|^2$,
and hence the calculated decay widths,
are $2\sim 3$ orders of magnitude smaller than needed for 
consistency with the data.
This is the main new observation of this work.
%Part of the reason for the small widths is simply the fact that the
%constituent mesons are in P--wave and that the wavefunction at  
%the origin vanishes, as explained in Section \ref{decwid}.  
To understand this we estimate the ratio which governs the
width of a P--wave molecule relative to that of an S--wave
molecule (Eqs. \ref{swav}-\ref{pwav})  
\begin{eqnarray} 
R = \frac{|\psi_P^\prime (0)|^2/m^2}{|\psi_S(0)|^2},
\end{eqnarray}
where the subscripts $P$ and $S$ stand for the P-- and S--waves 
respectively and $m$ is the mass of the molecular state 
(whose existence can be understood from dimensional analysis).
%, in fact, it would 
%automatically appear 
%as a kinetic factor in the decay rate calculation.
As a guide, we note that for a Coulomb potential  
$$\frac{\psi_P^\prime (0)}{\psi_S(0)}\propto \mu \alpha,$$  
where $\mu$ is the reduced mass of the system and $\alpha$ is the
Coulomb coupling.  
For the ``heavy--light'' $1^{-+}$ 
molecular states consisting of one heavy meson
and one light meson assumed in this work,  
the reduced mass is close to the mass of the light meson
$\pi$, which is much
smaller than the mass of $\rhohat(1400)$ or $\rhohat(1600)$
(which is close to the mass of the heavy meson). Thus there
is a suppression factor $(\mu/m)^2< 0.01$ in the ratio $R$. Since
the decay widths of $f_0(980)$ and $a_0(980)$ are proportional to 
$|\psi_S(0)|^2$, but the decay widths of $1^{-+}$ molecular
states (if they exist) are proportional to $|\psi_P^\prime  
(0)|^2/m^2$, one can easily see that a factor of $10^{-3} - 10^{-2}$ would 
suppress the $1^{-+}$ decays.
  
There is a close correspondence between the exchange mesons we used,
and those used in nucleon--nucleon interaction models, e.g. the
phenomenologically successful Bonn potential~\cite{mach}. The exchange 
mesons are the same except for two cases. 
Firstly, the
$\eta$ and $\eta'$ are not employed in the Bonn potential because
their couplings to the nucleon are small. (For our molecules, these
exchanges are not allowed by parity.) 
Secondly, the $f_0(980)$ and
$\phi$ are not considered in the Bonn potential because their large
$s\bar{s}$ content implies Okubo--Zweig--Iizuka (OZI) rule violating 
couplings with the nucleon.
For our calculations, these mesons have to be considered because the
$K\bar{K}$ molecule has OZI rule allowed couplings to $s\bar{s}$
exchange mesons. The $\pi\eta(1295)$ and $\pi\eta(1440)$ molecules
can also experience OZI rule allowed $f_0(980)$ exchange, because the 
$f_0(980)$ is not thought to be an almost pure $s\bar{s}$ state.
Unlike the Bonn potential, we do not consider higher order meson
exchange in this first orientation. 

In our derivations and numerical computations, there are 
some uncertainties from both experimental and theoretical sides. 
For example, the mass of $\sigma-$meson ranges from 400 to 
1200 MeV and its decay width is also uncertain. 
The model has obvious flaws due to introduction of
phenomenological parameters and ignoring the processes at
quark--level. Therefore
we can never expect very accurate predictions within this framework. If an 
additional factor of $2\sim3$ exists in the results, it is not surprising. 
However, since all derivations are based on the quantum field 
theory in a well motivated approximation,
and parameters are fixed by fitting data, order of
magnitude estimates should be valid.
The only conclusion we can expect is
whether the present data can tolerate the molecular state scenario 
for $\rhohat(1400)$ and $\rhohat(1600)$ within experimental errors.

Since the dissociation rates of the $1^{-+}$ molecular states calculated in
this work are two to three orders of magnitude smaller than the data,  
the lifetimes of the molecular states are fully determined by the
lifetimes of the constituent mesons ($\pi$, $\eta(1295)$ and
$\eta(1440)$ resonances). They are not comparable
to the measured values of the $\rhohat(1400)$ and $\rhohat(1600)$ lifetimes. 
%Moreover, the estimated relative ratios of $\rhohat(1600)\rightarrow  
%\pi b_1:\pi f_1:\pi\rho\approx 2.6:2:1$, obviously deviate from  
%that 170:60:10, which are evaluated in terms of the hybrid  
%assumption. 
So we conclude that the possibility of
$\rhohat(1400)$ and $\rhohat(1600)$ respectively being 
loosely bound $\pi\eta(1295)$ or  
$\pi\eta(1440)$ molecular states is 
ruled out by the data.
%, thus they would be favorable candidates of hybrid or glueball.

The decay formalism shows that the dissociation rate of a 
heavy--light molecule
with a binding energy much smaller than the masses of the constituent
mesons becomes smaller the higher the relative angular momentum between the
constituent mesons is.
The explicit calculation suggests that the dissociation rate is already
negligible for a P--wave molecule. We hence expect the lifetime
of a molecule with one or more unit of relative angular momentum 
to be fully determined by the lifetimes of its constituent mesons.
Contrary to the usual expectation for molecules~\cite{ted}, such a
molecule will have 
negligible ``fall--apart'' decay by dissociation to the two mesons that it 
is composed of. Instead, it will decay via the decay of its constituent 
mesons.
A constituent meson will emit at least two particles in its decay
implying that the decay of the molecule is dominated by its decay
to three or more particles. This is an unusual decay pattern.

If an experimental state is found with a lifetime
equal to the lifetimes of its constituents, a probable explanation is that
it is a molecule which has a sufficiently low binding energy so that 
decay due to dissociation is small. If the constituent mesons need to be  
in P--wave to obtain the desired quantum numbers of the state, one would
usually expect a lower--lying S--wave state with different quantum numbers
and an increased dissociation rate.
This could make the S--wave state so broad that it would be
difficult to observe. We hence predict the 
possibility of longer living orbitally excited molecular states.
  
%The molecular hypothesis need to be contrasted with the hypothesis that 
%either of the states are hybrid mesons. In the flux--tube model for
%hybrid mesons, there is a selection rule that suppresses decay to
%e.g. $\rho\pi$.  
%Close~\cite{Close} estimated
%that for I=1 states with mass around 2 GeV, partial widths are
%typically
%$$\pi b_1:\pi f_1:\pi\rho\approx 170:60:10\;{\rm MeV}.$$ 
%The absolute values seem too large for the lifetime of 
%$\rhohat(1600)$, but the ratios make sense and will be measured in future  
%experiments.
%As discussed in ref.~\cite{Close},  
%the large difference between $(b_1 , f_1)\;\pi$
%and $\rho\pi$ is due to the difference between the
%internal structure of $b_1,\; f_1$ versus the $\rho$.  

\vspace{.7cm}\noindent{\bf Acknowledgment}
  
This work is partially supported by the National Natural Science 
Foundation of China, and the U.S. Department of Energy under contract 
W-7405-ENG-36.
We would like to thank Y.-C. Zhu and B.-S. Zou for helpful discussions.
We would also like to thank the participants of the HADRON 99
conference where the topic of exotic mesons was addressed and we began  
to work on this issue. Moreover, we benefited from discussions with
G.-L. Wang who provided his expertise on the B--S approach.
  
\begin{appendix}

\section{Appendix: General formalism\label{app}}
 
For interactions among mesons, the early theory is the linear  
$\sigma$ model and the SU(2) Lagrangian is \cite{Don} 
\begin{equation}\label{sigeqn}
L={1\over 2}[(\partial_{\mu} S)^2-2m^2S^2]+\left({v+S\over     
2}\right)^2[(\partial_{\mu}\vec{\phi})^2+
{(\vec{\phi}\cdot\partial\vec{\phi})^2  
\over v^2-\vec{\phi}^2}]-\lambda vS^3-{\lambda\over 4}S^4+..., 
\end{equation}  
where   
$$S=\sqrt{\sigma^2+\vec{\pi}^2}-v,\;\;\; \vec{\phi}={v\vec{\pi}
\over\sqrt{\sigma^2+\vec{\pi}^2}},\;\;\; v=\sqrt{m^2\over\lambda},$$   
%and     
%$$\Sigma=\sigma+i{\bf \tau}\cdot{\bf \pi}=(v+S)U,;\;\;\; 
%U=exp(i{\bf \tau}\cdot{\bf \pi'},\;\;\; \pi'=\pi+....$$
and $\vec{\pi}$ and $\sigma$ are respectively the pion and $\sigma$--meson
fields.
     
Taking into account the interaction with vector mesons, one    
replaces the differentiation symbols with covariant   
differentiation forms. The effective chiral Lagrangian for
pseudoscalar and vector mesons is \cite{Wang}    
\begin{eqnarray} 
\label{L1}
L =\frac{F^2}{16}Tr(\nabla_\mu U \nabla^\mu U^\dag) +    
\frac{1}{4}m_0^2 Tr(V_\mu V^\mu),    
\end{eqnarray}   
where the vector meson fields are    
\begin{equation} 
V_\mu = \lambda^a V_\mu^a = \sqrt{2}\left(    
\begin{array}{ccc}     
  \frac{\rho^0_{\mu}}{\sqrt{2}}+  \frac{\omega_{\mu}}{\sqrt{2}} &     
\rho^+_{\mu} & K^{*+}_{\mu} \\
  \rho^-_{\mu} & -\frac{\rho^0_{\mu}}{\sqrt{2}}+  
\frac{\omega_{\mu}}{\sqrt{2}} & K^{*0}_{\mu} \\   
  K^{*-}_{\mu} & \bar{K}^{*0}_{\mu} & \phi_\mu    
\end{array} \right)    
\end{equation}   
with $\lambda^a\; (a=1,...,8)$ the SU(3) generators. 
The SU(3) differentiation forms are
\begin{eqnarray} 
\nabla_\mu U = \partial_\mu U-ir_\mu U +iUl_\mu , \nonumber\\    
\nabla_\mu U^\dag = \partial_\mu U^\dag -il_\mu U^\dag +iU^\dag r_\mu ,
\end{eqnarray}   
where $l_\mu=v_\mu+a_\mu$ and $r_\mu=v_\mu-a_\mu$ are linear     
combinations of the external vector $v_\mu$ and axial vector $a_\mu$
fields. Also
%, such as $\rho,\; \omega,\; K^*,\; \phi$ and $b_1,\; h_1,\; a_1$ etc.     
\begin{eqnarray} 
U = \exp(i\lambda^a \Phi^a),
\end{eqnarray}   
with $\Phi^a$ the fields of the pseudoscalar meson octet. 
    
From recent studies, the $\sigma-$meson may be an independent field,    
so that we retain the interaction in Eq. \ref{sigeqn}.    
In our final expression, we keep the leading $\sigma-$related terms and
the leading terms in the pseudoscalar--vector interactions,     
while we ignore all axial vectors because they are much    
heavier. We also extend the Lagrangian to include the iso--scalars     
$\eta(1295)$ and $\eta(1440)$. The much simplified Lagrangian
reads     
\begin{equation}\label{laf} 
L_{eff}=g\sigma PP+g_{ijk}'v^i_{\mu}P^j\partial^{\mu} P^k+..., 
\end{equation}   
where $i,j,k$ guarantees a proper SU(2) combination, and the ellipsis
denotes the smaller contributions which are neglected in our    
calculations. We note that the symmetry reduces from SU(3) to     
SU(2).
%, but two excited states $\eta(1295)$ and $\eta(1440)$ are involved.   

%The effective vertices between the mesons (pseudoscalar or vector) 
%are given in an effective chiral Lagrangian as 
%\begin{eqnarray} 
%L =\frac{F^2}{16}Tr(\nabla_\mu U \nabla^\mu U^\dag) +  
%\frac{1}{4}m_0^2 Tr(V_\mu V^\mu) 
%\end{eqnarray}
%where  
%\begin{equation} 
%V_\mu(x) = \lambda\cdot V_\mu = \sqrt{2}\left( 
%\begin{array}{ccc} 
%  \frac{\rho^0_{\mu}}{\sqrt{2}}+  \frac{\omega_{\mu}}{\sqrt{2}} &
%\rho^+_{\mu} & K^{*+}_{\mu} \\ 
%  \rho^-_{\mu} & -\frac{\rho^0_{\mu}}{\sqrt{2}}+  
%\frac{\omega_{\mu}}{\sqrt{2}} & K^{*0}_{\mu} \\
%  K^{*-}_{\mu} & K^{*0}_{\mu} & \phi_\mu
%\end{array} \right)
%\end{equation}
  
%In the expression
%\begin{eqnarray} 
%\nabla_\mu U = \partial_\mu U-i r_\mu U +iUl_\mu , \nonumber\\ 
%\nabla_\mu U^\dag = \partial_\mu U^\dag -il_\mu U^\dag +iU^\dag r_\mu , 
%\end{eqnarray}
%where $l_\mu=v_\mu+a_\mu$ and $r_\mu=v_\mu-a_\mu$ are linear
%combinations of the external vector and axial vector fields.
%\begin{eqnarray} 
%U(\Phi) = exp(i\lambda^a \Phi^a(x)),
%\end{eqnarray}
%and $\lambda^a(a=1,...,8)$ are the SU(3) generators.
  
\end{appendix}

\end{document}